\documentclass[A4,twocolumn]{webofc}
\usepackage[varg]{txfonts}   
\woctitle{?????????}
\begin{document}
\title{Low-Energy Magnetic Radiation}\footnote{Part of the material and additional  material is published in \cite{Schwengner14,Frauendorf14}.}
%
%

\author{S.Frauendorf\inst{1}\fnsep\thanks{\email{sfrauend@nd.edu}} \and
        M. Beard\inst{1} \and
        M Mumpower\inst{1} \and
        R. Schwengner\inst{2} \and
        K. Wimmer\inst{3}
}

\institute{University Notre Dame, IN 46557, USA 
\and
           IRP, HZDR, 01328 Dresden, Germany
\and
           Central Michigan University,  Mt. Pleasant, MI 48859, USA 
          }

\abstract{%
  A pronounced spike at low energy in the strength function for magnetic radiation (LEMAR) is found by means of Shell Model calculations,
 which explains the experimentally observed enhancement  of the dipole strength. 
 LEMAR originates from statistical low-energy M1-transitions between many excited complex states.
  Re-coupling of the proton and neutron high-j orbitals generates the strong magnetic radiation. 
  LEMAR is predicted for nuclides with $A\approx 132$ participating in the r-process of element synthesis.
 It increases  the reaction rates by a factor of 2.5. The spectral function of LEMAR follows Planck's Law. A power law for the
 size distribution of the $B(M1)$ values are found. 
}
\maketitle
\section{Introduction}
\label{intro}
\begin{figure}[h]
\centering
\includegraphics[width=0.9\columnwidth]{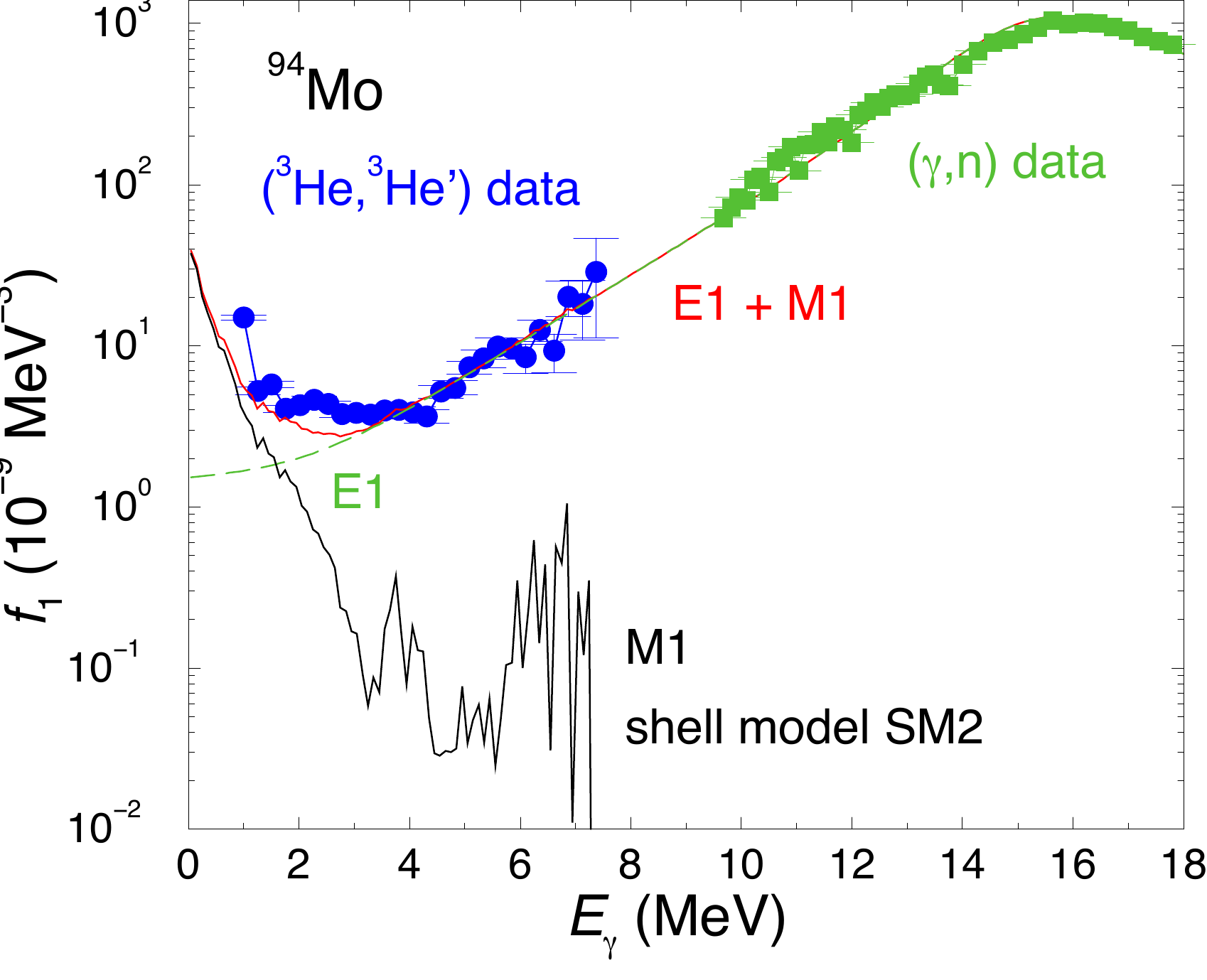}
\caption{ Strength functions for $^{94}$Mo 
deduced from ($^3$He,$^3$He') (blue circles) and $(\gamma,n)$ (green squares)
experiments, the $M1$ strength function from the present Shell Model
calculations (black solid line), $E1$ strength according to the GLO analytical
expression (green dashed line), and the total ($E1 + M1$) dipole strength
function (red line). }
\label{fig:94MoEgf1}      
\end{figure}

 Photonuclear reactions and the inverse radiative-capture reactions between
nuclear states in the region of high excitation energy and large level density
are of considerable interest in many
applications. 
A critical input to
calculations of the reaction rates is the average strength of the cascade of 
$\gamma$-transitions  de-exciting the nucleus,
which is described by photon strength function.
 An increase of the
dipole strength function below 3 MeV toward low $\gamma$-ray energy has recently been
observed  in  nuclides in the mass range from $A \approx$ 40 to 100, in particular, 
using ($^3$He,$^3$He') reactions on various Mo isotopes \cite{gut05}. 
The data are shown in
Fig.~\ref{fig:94MoEgf1}. Around \mbox{1 MeV}, the experimental strength function (blue) is about a factor of 10 
larger than expected for a damped Giant Dipole Resonance shown by the dashed green curve, which 
is calculated by the standard GLO expression commonly used for describing the strength of 
electric dipole (E1) radiation  in this energy region.

The enhancement is not observed in the inverse process of absorbing $\gamma$-quanta
by nuclei in the ground state. 
Only few discrete lines are found within the interval of the first 4 MeV  \cite{Pietralla99}. 
The enhancement in the de-excitation cascade must be related to the complex structure of the highly
excited states among which the transitions occur. We plotted the reduced
probabilities of all  discrete transitions reported for the nuclides  with $88\leq A \leq 98$ depending on their 
transition energy. The reduced probabilities of the magnetic transitions $B(M1)$ clearly increase toward zero transition energy,
whereas no such tendency is seen for reduced probabilities $B(E1)$ for the electric transitions. Based on this observation we 
conjectured that the enhancement seen in experiments like the one in Fig. \ref{fig:94MoEgf1} is caused by $M1$ transitions between
high-lying states.  To study this conjecture, we carried out Shell Model calculations for 
the nuclides $^{94,95,96}$Mo and $^{90}$Zr, for which the enhancement has been observed in experiment.

\section{Shell Model Calculations}

The Shell Model calculations  included the active proton orbits
$\pi(0f_{5/2}, 1p_{3/2}, 1p_{1/2}, 0g_{9/2})$ and the neutron orbits
 $\nu(0g_{9/2}, 1d_{5/2}, 0g_{7/2})$ relative to a $^{68}$Ni core. 
The set of empirical matrix elements for the effective
interaction and of the single particle energies are
 given in  Refs. \cite{Schwengner14,sch09}. 
  For calculating the  reduced transition probabilities $B(M1)$  effective
$g$-factors of $g^{\rm eff}_s = 0.7  g^{\rm free}_s$ have been applied.

To make the calculations feasible truncations of the occupation numbers were applied.
Up to two protons could be lifted from the $1p_{1/2}$ orbit to the $0g_{9/2}$ orbit.
In $^{94,95}$Mo, one neutron from the $0g_{9/2}$ orbit and one neutron from $1d_{5/2}$ orbit
may be excited to the $0g_{7/2}$ orbit or one neutron may be lifted from the $0g_{9/2}$ orbit
to the $1d_{5/2}$ orbit and one from the $1d_{5/2}$ orbit to the $0g_{7/2}$ orbit.
In $^{90}$Zr, one neutron from the $0g_{9/2}$ orbit may be excited to the
$1d_{5/2}$ orbit and one to the $0g_{7/2}$ orbit or two neutrons from the $0g_{9/2}$ orbit
may be excited to the $0g_{7/2}$ orbit.

\begin{figure}[h]
\includegraphics[width=\columnwidth]{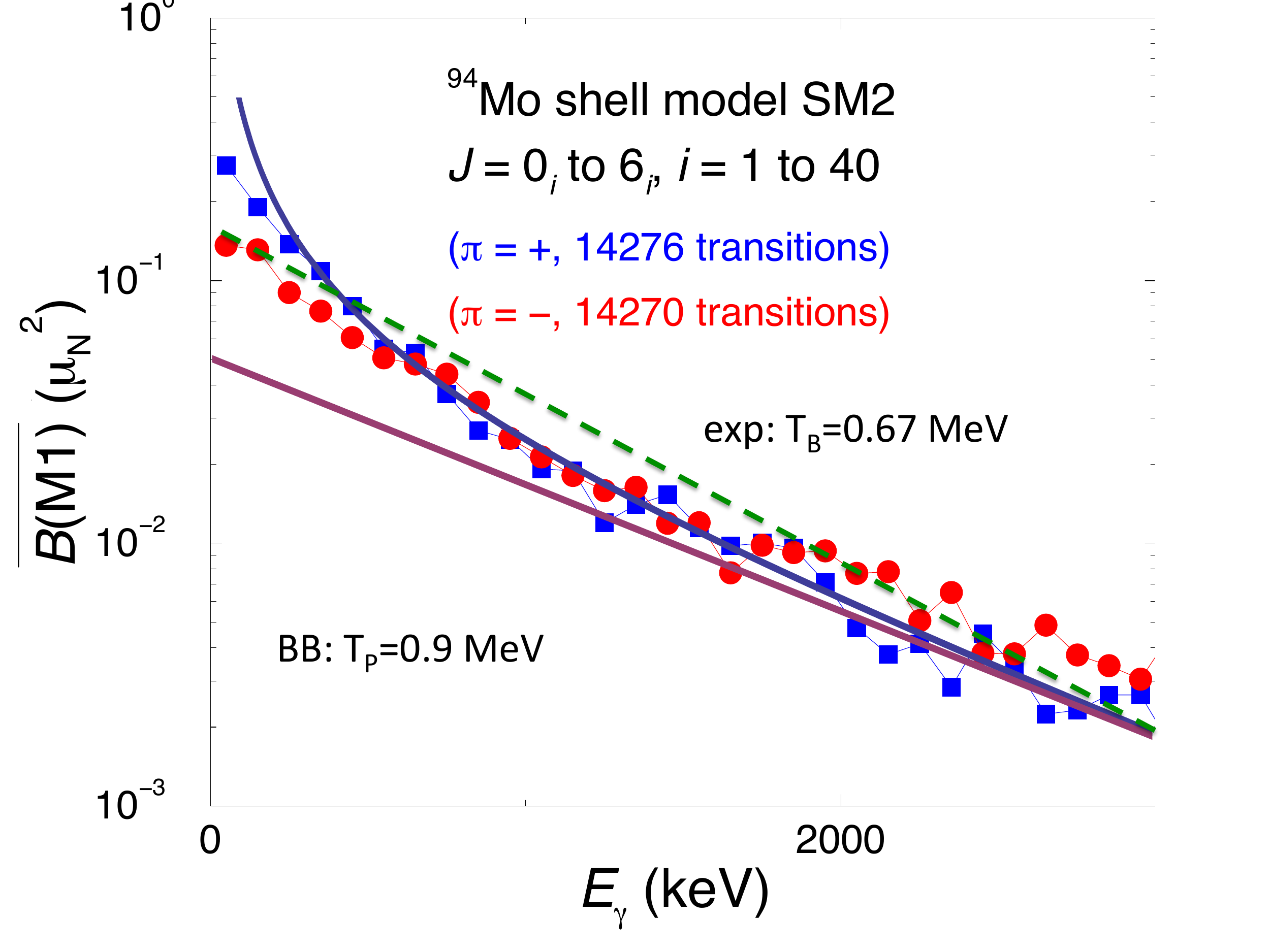}
\caption{ \label{fig:94MoEgM1} Average $B(M1)$ values in 100 keV
bins of the transition energy calculated for positive-parity (blue squares) and
negative-parity (red circles) states in $^{94}$Mo. The dashed line (exp) shows  
Eq. (\ref{eq:BM1exp}) with $T_B=0.67$ MeV.
 The curve (BB) shows the Black Body distribution (\ref{eq:BB}) with
 the temperature $T_P=0.9$ MeV, where
the straight line is the high-energy limit $B_P\exp(-E\gamma/T_P)$.}
 \end{figure}

The calculations included states with spins from $J$ = 0 to 6 for $^{90}$Zr and
$^{94}$Mo and from $J$ = 1/2 to 13/2 for $^{95}$Mo. For each spin the lowest 40
states were calculated. The reduced transition probabilities $B(M1)$  were
calculated for all transitions from initial to final states with energies
$E_f < E_i$ and spins $J_f = J_i, J_i \pm 1$. For the minimum and maximum
$J_i$, the cases $J_f = J_i - 1$ and $J_f = J_i + 1$, respectively, were
excluded. This resulted in more than 14000 $M1$ transitions for each parity
$\pi = +$ and $\pi = -$, which were sorted into 100 keV bins according to
their transition energy $E_\gamma = E_i - E_f$. The average $\overline{B}(M1,E_\gamma)$ value for
one energy bin was obtained as the sum of all $B(M1)$ values divided by the
number of transitions within this bin. The results for 
$^{94}$Mo are shown in Fig.~\ref{fig:94MoEgM1}.

Fig.~\ref{fig:94MoEgM1} demonstrates that,
up to 2 MeV, the LEMAR spike of $\overline{B}(M1,E_\gamma)$ is 
approximated by the exponential function
\begin{equation}\label{eq:BM1exp}
\overline{B}(M1,E_\gamma) = B_0 \exp{(-E_\gamma/T_B)},
\end{equation}
 with $B_{0} = \overline{B}(M1,0)$ and $T_B$ being constants. This is the case for all studied cases.

The  exponential dependence on the transition energy is retained by
the $M1$ strength functions, which are defined by the relation 
\begin{equation}
f_{M1}(E_\gamma)
= 16\pi/9 (\hbar c)^{-3}\overline{B}(M1,E_\gamma)\rho(E_i),
\end{equation}
where the level density at the initial state  $\rho(E_i)$ is obtained from the Shell Model calculations.
 The level densities
$\rho(E_i,\pi)$ were determined by counting the calculated levels within energy
intervals of 1 MeV for the two parities separately. 
These combinatorial level densities were used in calculating the strength functions
 by means of Eq. (2). As seen  in Fig. \ref{fig:94MoEgf1},  there is a pronounced enhancement below 2 MeV, which is
well described by the exponential function
\begin{equation}
f_{M1}(E_\gamma) = f_0 \exp{(-E_\gamma/T_f)}.
\end{equation}
For $^{90}$Zr, $^{94}$Mo, $^{95}$Mo, and $^{96}$Mo, the parameters are
$f_0$ = (34, 37, 39, 55) $\times$ 10$^{-9}$ MeV$^{-3}$ and 
$T_f$ = (0.50, 0.50, 0.51, 0.48) MeV, respectively.
The calculated M1- enhancement is consistent with the experiment, 
which however did not determine wether the radiation is electric or magnetic.

In the case of the Mo isotopes, the LEMAR transitions mainly originate 
from states between 2 and 4 MeV with a tail extending to 6 MeV.
 In $^{90}$Zr, the distribution starts at about 3 MeV and continues to 10 MeV.
Within the studied range of $0 \leq J\leq 6$, the transitions originate
with about the same probability  
from initial states with different angular momentum. 


{\bf LEMAR is generated by a huge number of weak low-energy  $M1$ transitions, which originate from high-lying states
and add up to strong $M1$ radiation. LEMAR accounts for the observed low-energy enhancement of the strength function.
}

\section{Origin of the M1-strength}

LEMAR is caused by transitions between
many close-lying states of all considered spins located well above the yrast
line in the transitional region to the quasi-continuum of nuclear states.  
Inspecting the composition of initial and final states, one finds large $B(M1)$ values for transitions
between states that contain a large component (up to about 50\%) of the same
configuration with broken pairs of both protons and neutrons in high-$j$
orbits. The largest $M1$ matrix elements connect configurations with the spins
of high-$j$ protons re-coupled with respect to those of high-$j$ neutrons to
the total spin $J_f = J_i, J_i \pm 1$. The main configurations are
$\pi(0g_{9/2}^2) \nu(1d_{5/2}^2)$, 
$\pi(0g_{9/2}^2) \nu(1d_{5/2}^1 0g_{7/2}^1)$, and 
$\pi(0g_{9/2}^2) \nu(1d_{5/2}^2 0g_{9/2}^{-1} 0g_{7/2}^1)$ for positive-parity
states in $^{94}$Mo. Negative-parity states contain a proton lifted from the
$1p_{1/2}$ to the $0g_{9/2}$ orbit in addition. The orbits in these configurations have large 
$g$-factors with opposite signs for protons and neutrons. Combined with specific
relative phases of the proton and neutron partitions they cause large total
magnetic moments. 
The residual interaction between the valence particles and holes generates an energy difference between the states 
related to each other by recouping the angular momenta, which were degenerated without it. These energetic splittings enable 
transitions between the states by emitting an $M1$ $\gamma$-quant. In this sense, it is the residual interaction that generates the radiation.


Magnetic Rotation (MR)  \cite{fraRMP} is a known example of the re-coupling mechanism
 generating
strong low-energy magnetic transitions between near-yrast states.
The configurations are rather pure, and the transition energies increase with angular momentum forming regular rotational sequences.
The high-lying states that generate the LEMAR
 are composed of a strong mix of configurations. The complex mixing changes the residual interaction   between
the states. As a  consequence,  the distance between the states becomes randomized. 
Because there are several high-j orbitals involved in generating LEMAR there are other re-coupling 
possibilities leading to strong M1 transition than the simple type generating MR.
  
{\bf LEMAR consists of transitions between states related by angular momentum re-coupling
of the same high-j proton and neutron configurations.  Transition energies and probabilities are randomized.
LEMAR is closely related to Magnetic Rotation.
}
\begin{figure}[h]
\includegraphics[width=0.7\columnwidth,angle=-90]{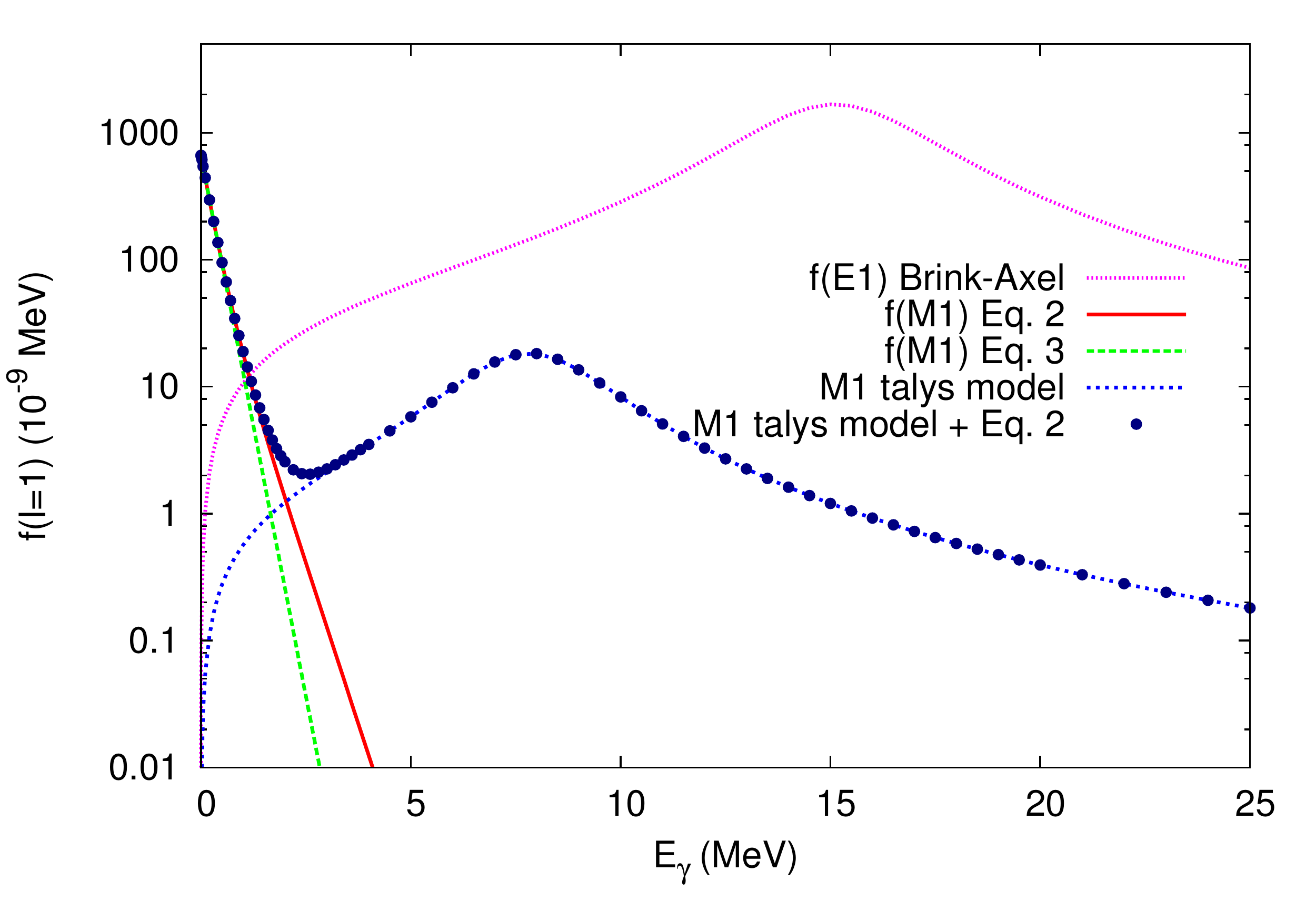}
\caption{\label{fig:131Cdf1}The $\gamma$ - strength function used in the calculation of the $^{130}Cd(n,\gamma)$ reaction rate. 
The standard M1 and E1 strength functions are denoted by "talys model" and "Brink-Axel", respectively.  "M1  Eq. 2" and "M1 Eq. 3" refer to calculating 
the LEMAR contribution by means of Eqs. (2) and (3), respectively.  }
\end{figure}


\section{Consequences for the astrophysical r-process}

LEMAR is expected to enhance the rate for the $(n,\gamma)$ reactions compared with standard calculations, an example of which 
is illustrated Fig. \ref{fig:131Cdf1}. The region with proton number below $Z=50$ and 
neutron number above $N=82$ plays a key role in the r-process of element synthesis in violent stellar events.
For this reason we studied $^{131}$Cd, which is a waiting point in the reaction chain. The Shell Model calculation 
was performed within the model space of the  $1p_{3/2}$, $0f_{5/2}$, $1_{p1/2}$, $0g_{9/2}$   proton holes and
 $0h_{9/2}$, $1f_{7/2}$, $1f_{5/2}$, $2p_{3/2}$, $2p_{1/2}$ neutrons particles using a 
  G-matrix derived from the CD-Bonn NN interaction. Fig. \ref{fig:131Cdf1} shows that the LEMAR spike is quite similar 
  to the ones for the stable Mo isotopes.
  Including LEMAR into  the M1-strength function model used in the statistical model code TALYS 
  code increases the M1 strength function by approximately five orders of magnitude at very low energy. 
  As a result, the M1 strength function dominates the E1 for $E_{\gamma}$ < 2~MeV. 
  Adding LEMAR to the standard M1-strength function, a calculation by means of the  TALYS   increases the rate 
  of the $^{130}Cd(n,\gamma)$ reaction rate by a factor of 2.5 over a wide temperature range of the stellar environment.

\begin{figure}[h]
\includegraphics[width=\columnwidth]{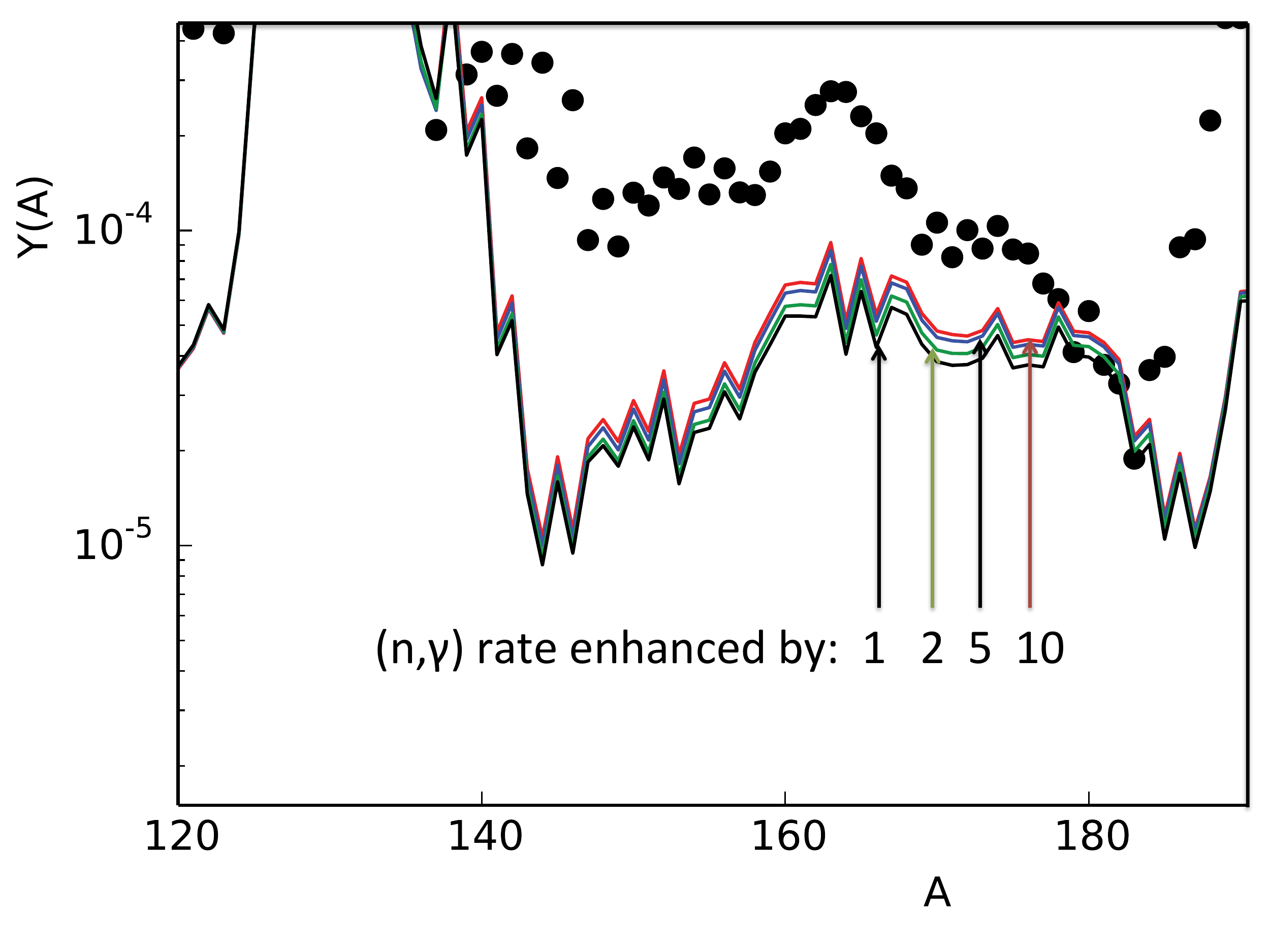}
\vspace*{-0.9cm}
\caption{\label{fig:cold-r} An estimation of the impact of LEMAR on cold r-process abundance predictions. 
Neutron capture rates of nuclei in the region defined by $N=82$ to $N=88$ and $Z=45$ to $Z=50$ were enhanced by constant factors of 2, 5 and 10. Solar r-process residuals (black dots) from \cite{Arlandini1999}.) }
\end{figure}

Investigations of LEMAR remain on going for additional nuclei in this and other regions that are applicable to r-process nucleosynthesis. 
The current limiting factor is the computational cost of Shell Model calculations. 
To estimate the impact of LEMAR on r-process abundance predictions we have artificially enhanced rates of nuclei in the region defined by $N=82$ to $N=88$ and $Z=45$ to $Z=50$. 
Fig. \ref{fig:cold-r} shows the results the context of a `cold' r-process, where  
equilibrium between neutron captures and photo-dissociations is short-lived or nonexistent. 
After nuclei fall out of $(n,\gamma)\rightleftarrows(\gamma,n)$ equilibrium individual neutron capture rates have the capacity to impact the predicted abundances. 

{\bf The neutron capture rate enhancement from LEMAR can potentially boost final abundance predictions after the second ($A\sim130$) peak which partially fills in the deficiencies found in this sector relative to the baseline calculation where the rates remained unchanged. 
This abundance boost better matches the solar isotopic r-process residuals (black dots). }

\section{Statistical Properties of the Transitions}
As the LEMAR is generated by a huge number of weak transitions 
between complex  states, it is natural to study  their statistical
characteristics.  The study is still on the way, and we report only tentative results
obtained so far.

As illustrated by  Fig. \ref{fig:94MoEgM1} the average reduced transition probabilities $\overline{B}(M1,E_\gamma)$ decrease approximately exponentially with the
 energy difference between initial and final states of the transitions. The decrease is determined by the 
 parameter $T_B$ in Eq. (1), which scatters around 0.5 MeV for the studied nuclides.
 The strength functions decrease exponentially 
 as well, with the characteristic parameter in Eq. (3)  $T_f\approx 0.5$ MeV.  The mean value of the transition energy for 
 the strength function is $T_f\approx 0.5$ MeV. 
 
 However as seen in  Fig. \ref{fig:94MoEgM1}, there is a slight convex curvature, which appears systematically
 for all studied nuclei.  The curvature is well accounted for by the modified expression 
 \begin{equation}\label{eq:BB}
\overline{B}(M1,E_\gamma) = \frac{B_P}{ \exp{(E_\gamma/T_P)-1}},
\end{equation}
 with $B_{P} $ and $T_P$ being constants. This is the case for all studied cases.
Multiplying the expression by $E_\gamma^3$, the phase space factor of dipole photons, one obtains the spectral function of the 
LEMAR radiation width
 \begin{equation}\label{eq:Planck}
\Gamma(E_\gamma) = \frac{\Gamma_P\left(E_\gamma/T_P\right)^3}{ \exp{(E_\gamma/T_P)-1}},
\end{equation}
with $B_{P} $ and $T_P$ being constants. {\bf The expression is recognized as  Planck's formula for Black Body radiation!}
 
\begin{figure}[h]
\includegraphics[width=\columnwidth]{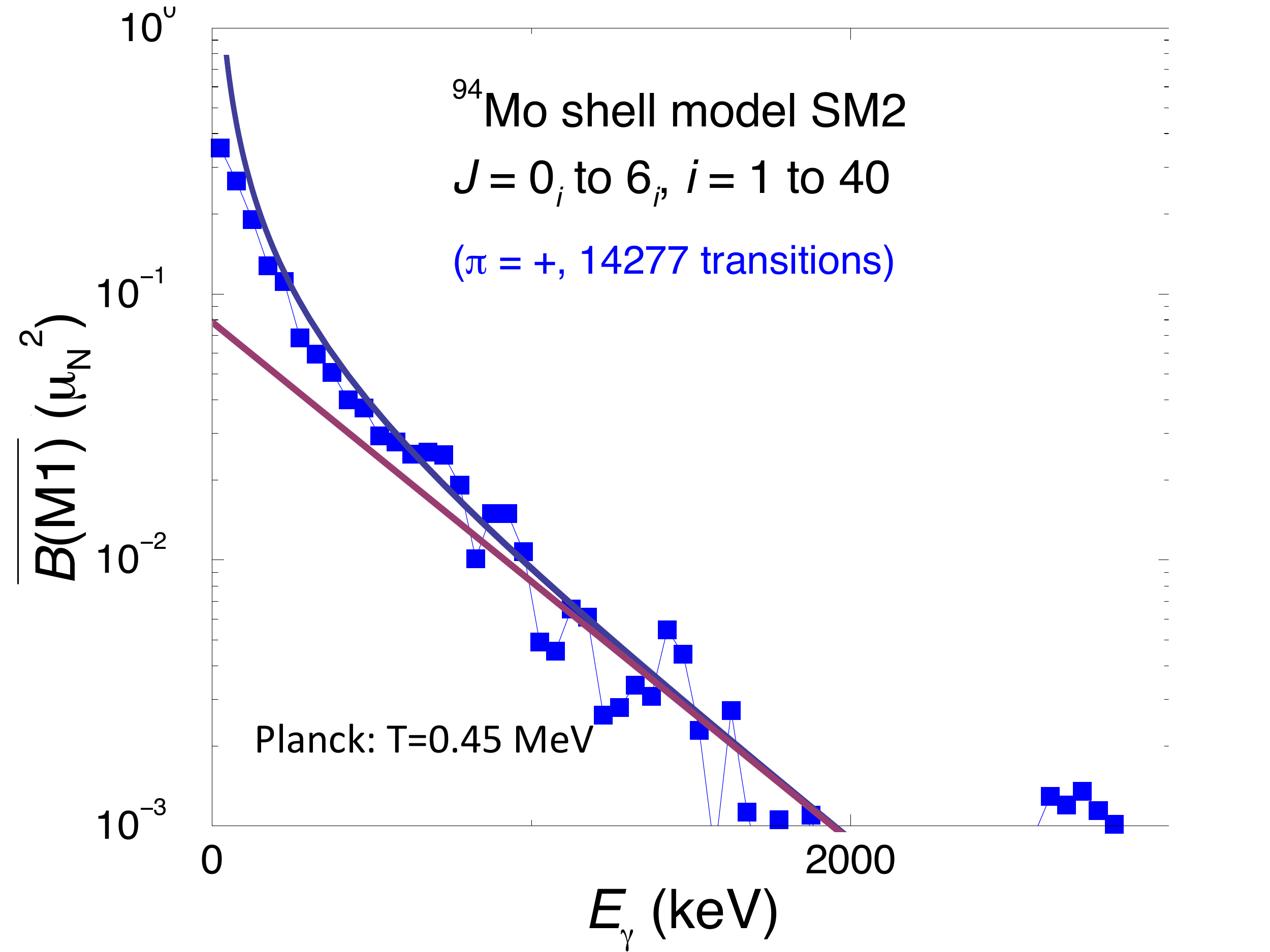}
\caption{\label{fig:BM1logMo94fhalf} Average $B(M1)$ values in 100 keV
bins of the transition energy calculated for positive-parity states in $^{94}$Mo calculated with the half strength of the interaction.
 The curve shows Planck's law with the temperature $T_P=0.45$ MeV.
For comparison, the straight line shows the high-energy limit $\Gamma_0\exp(-E\gamma/T_P)$.  }
\end{figure}

The appearance of the Black Body (BB)  spectral function  
  is new for nuclei.  Familiar  is a dependence on the transition energy corresponding to a Lorentzian   resonance 
 caused by a doorway state. Examples are  the Giant Dipole Resonance in the $E1$ strength function and the spin-flip
 resonance in the $M1$ strength function shown in Fig. \ref{fig:131Cdf1}. The resonance energy 
 is a combination of a characteristic particle-hole energy with the residual interaction, and the width reflects 
 the coupling to the background of complex states.  A BB spectrum appears when there is no such characteristic frequency,
 and the radiating constituents freely respond to the thermal activation. Examples are the free electrons in metals or the electrons and protons in
 the solar plasma. The spectra of hot metals and of the sun are almost perfectly BB. In the case of LEMAR, the mutual reorientation of the high-j
 orbitals in the spherical mean field does not cost energy, the residual interaction acts in a random way as a thermal activation, and a BB spectrum is the result.
 Fig. \ref{fig:BM1logMo94fhalf}  substantiates the point. It shows a a Shell Model calculation with all interaction matrix elements multiplied by a factor of 0.5.
 Clearly, the distribution is well approximated by  the BB formula (\ref{eq:BB}) with half the temperature of the calculation with the full interaction strength shown in  
 Fig. \ref{fig:94MoEgM1}. The residual interaction plays the role of a thermal agitation of the spins of the high-j orbitals, and the interaction strength 
 sets the agitation  scale, which is the temperature. 
\begin{figure}[h]
\includegraphics[width=\columnwidth]{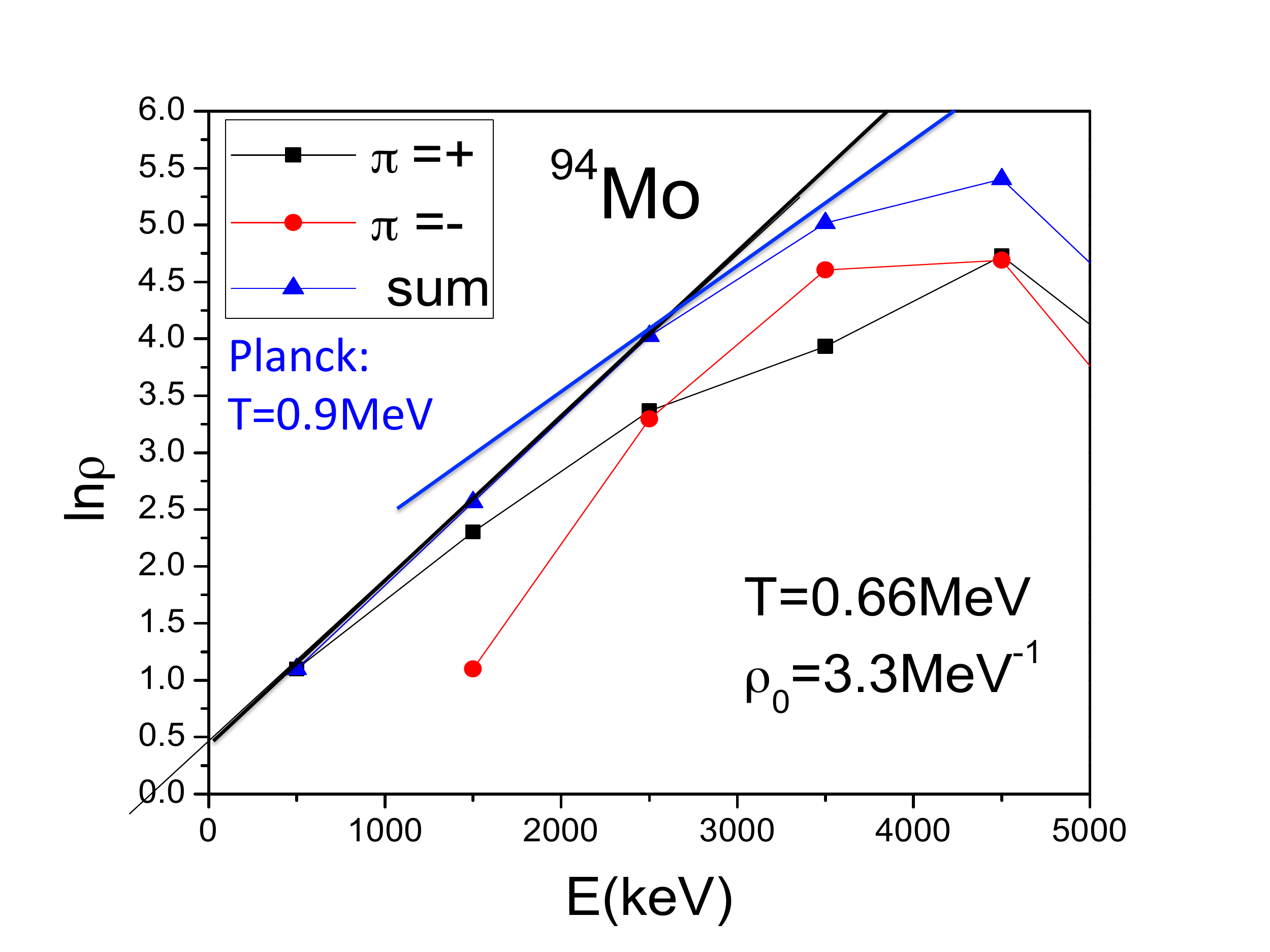}
\caption{\label{fig:densMo94} Level density of $^{94}$Mo calculated as the number of levels within bins of 1 MeV. The two straight lines  show the 
micro canonical temperature $1/T=dS/dE=d\ln(\rho)/dE$, where the the black line corresponds to $T=0.66$ MeV and the blue line to $T=0.9$MeV derived from the 
Planck distribution of the $\overline B(M1)$ values in fig. \ref{fig:94MoEgM1}.  }
\end{figure}

Fig. \ref{fig:densMo94} shows that the total level density $\rho(E)$ is well reproduced by the
constant-temperature (CT) expression
\begin{equation}\label{eq:rhoCT}
 \rho(E) = \rho_0 \exp{(E/T_\rho)}
 \end{equation}
with $T_\rho=0.66$ MeV  as long as $E <$ 2.5 MeV. 
The experimental level density,  as published in the RIPL3 data base \cite{RIPL3}, is very well described
 by  the CT expression (\ref{eq:rhoCT}) with $T_\rho=0.63$ MeV up to $E=3.8$ MeV, where the experimental
level scheme becomes incomplete. The Shell Model rather well reproduces the CT level density up to $E=2.5$ MeV.
For higher energies the combinatorial level density deviates
from the CT expression (\ref{eq:rhoCT}) and eventually decreases with excitation energy. Obviously, this due to
progressively missing  configurations   the present  Shell Model space  when $E$ increases beyond 2.5 MeV.

The value $T_\rho =0.66$ MeV of the 
constant temperature part of the combinatorial level density ($E<2.5$ MeV)
is smaller than the value $T_P$=0.9 MeV obtained from the BB distribution of the $\overline B(M1)$.
The difference may be explained by the fact that most of the transitions originate from states with $E_i>2.5$ MeV
where the entropy $S=\ln(\rho)$  deviates from the straight line  due to missing configurations in the calculation.  
The micro canonical temperature $T_\rho=\left(dS/dE\right)^{-1}$ increases with $E$ in this region,
and  the higher value of 
 $T_P$ arises as  some average over the region.  Accordingly,
the value of $1/T_P$ corresponds to the slope of $S$ near $E = 3.5$ MeV.  
Thus, the difference between $T_P$ and $T_\rho$ seems to be an artifact of Shell Model calculation.
$^{94}$Mo is representative for  the remaining nuclei studied. 
In the light of these findings we conjecture that $T_P=T_\rho$, i. e. that LEMAR is thermal radiation 
with a temperature that is equal to the micro canonical temperature of the level density. 
The experimental level densities are known to be very well accounted for by the CT expression (\ref{eq:rhoCT})
up to the neutron binding energies \cite{RIPL3}.  Thus one expects that the temperature is constant
over the range of excitation energies from which LEMAR is emitted. The experimental value for $^{94}$Mo, 
 $T=0.63$ MeV, should determine the spectral function of LEMAR in this nucleus.    If LEMAR spectrum  could be 
 sufficiently accurately measured, it could serve as a new thermometer of the nuclear temperature. 

\begin{figure}[h]
\includegraphics[width=\columnwidth]{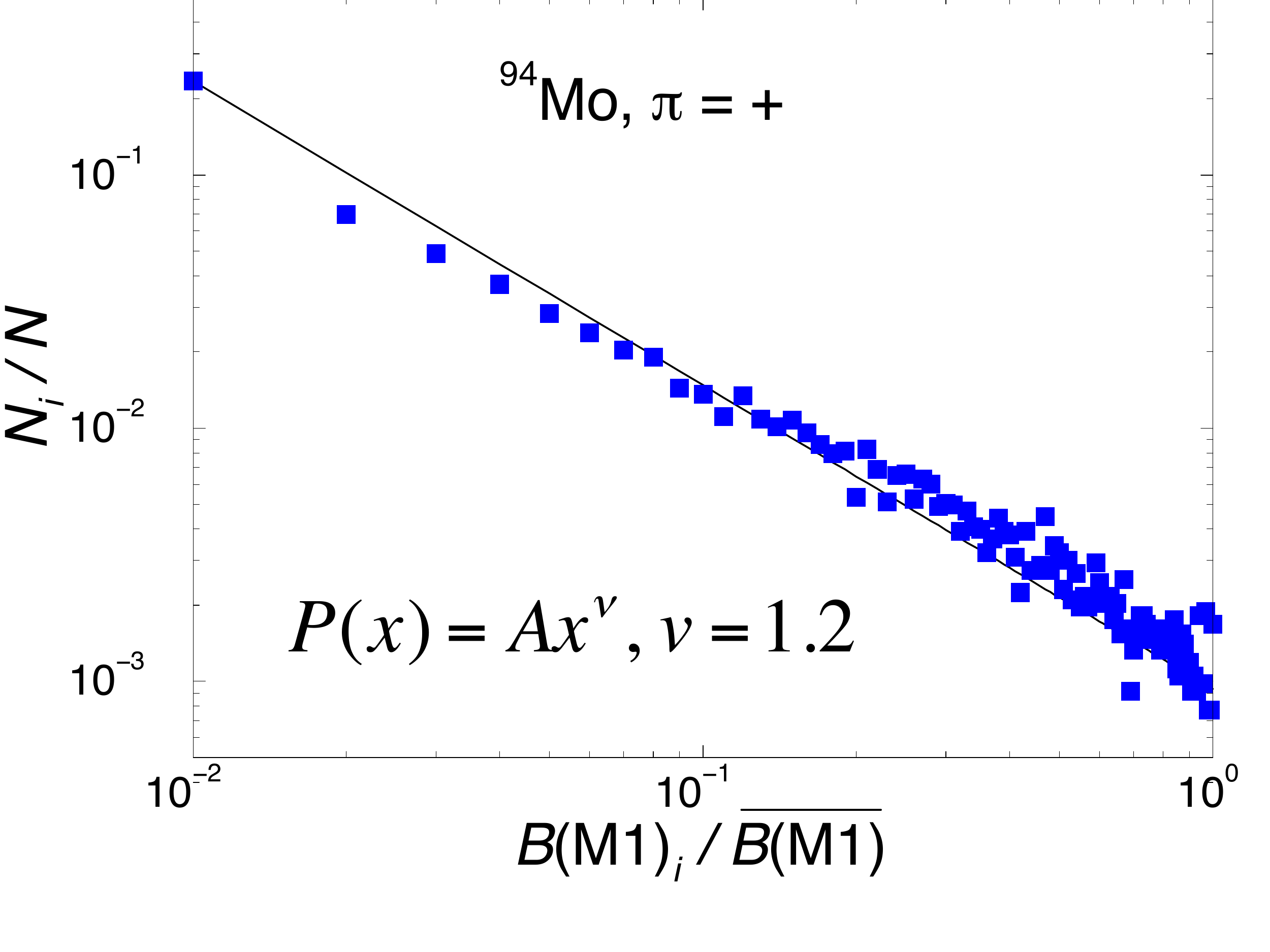}
\caption{\label{fig:NoPT} Probability distribution of the $B(M1)$ values in $^{94}$Mo for positive parity states
compared with a power law distribution (straight line ).}
\end{figure}

The size distribution of the $B(M1)$ values is shown in Fig. \ref{fig:NoPT},
where, $\overline{B(M1)}$ is the mean value over the complete distribution.
It follows the power law
\begin{equation}
 P(y) = Ay^\nu ~~~ y=B(M1)/\overline{B(M1)},
 \end{equation}  
 with $\nu=1.2$.
The distributions for  both parities in all three nuclides follow a power law with
the exponents $\nu$ scattering around 1.2.

The fact that the LEMAR distributions can be described by a simple power law is remarkable. 
Based on   Shell Model  \cite{ham02} and experimental \cite{shr00} 
studies of $\gamma$- transitions between complex 
excited states one rather would expect a Porter-Thomas-like distribution.
Power-law distributions are characteristic for scale-free 
systems, as for example the distribution of the number of clicks/site in the internet (Any site can connect with any number of other sites.)
or the  heat capacity near a second order phase transition (There is no scale for the fluctuations of the order parameter.).
LEMAR may be classified as  scale-free as follows. The various configurations that are related by re-coupling of the angular momentum
have all the same energy. The energy difference between the mixed states is generated by the residual interaction, which acts in a random 
way between the complex states. This differs from the conventional situation, where the single particle level spacing   
represents an energy scale for the various configurations mixed by the residual interaction. 

Statistical self-similarity is another  signature 
of scale-free systems. It means that the statistical characteristics of the system are the same when studied on different  scales
(The length of coast lines is a popular example.). LEMAR exhibits statistical self-similarity.  Fig. \ref{fig:BM1logMo94fhalf} shows the $\overline B(M1)$
values calculated by multiplying  residual interaction by a factor of 0.5 (i. e. changing the scale).  Comparing  with the 
results for the full interaction in Fig. \ref{fig:BM1logMo94full} , it is seen that the distribution 
remains BB, where the characteristic parameter $T_P=0.45$ MeV is one half of $T_B=0.9$ MeV of the  full calculation.    
 Remarkably, the condition for the appearance of the thermal black-body spectrum (no characteristic frequency) 
agrees with the prerequisite for a power-law distribution (scale-freeness).

{\bf LEMAR is thermal radiation. Its spectral function follows Planck's Law with a temperature $T_P$.
The value of $T_P$ seems to agrees with the micro canonical temperature $T_\rho$ of  combinatorial level density.  
The size distribution of the reduced transition probabilities is a power law with an exponent of $\sim 1.2$.
Both Planck's Law and the power law seem to originate from the absence of an energy scale for the LEMAR transitions.
}

 {\bf Acknowledgement:} Supported by the Grant No. DE-FG02-95ER4093 of the US Department of Energy.

\end{document}